\documentclass[english]{IEEEtran}

\pdfoutput=1

\usepackage[T1]{fontenc}
\usepackage[latin9]{inputenc}
\usepackage{verbatim}
\usepackage{float}
\usepackage{amsthm}
\usepackage{amsmath}
\usepackage{amssymb}
\usepackage{graphicx}
\usepackage{color}
\usepackage{url}
\usepackage{cite}
\makeatletter

\floatstyle{ruled}
\newfloat{algorithm}{tbp}{loa}
\providecommand{\algorithmname}{Algorithm}
\floatname{algorithm}{\protect\algorithmname}

\numberwithin{equation}{section}
\numberwithin{figure}{section}
\theoremstyle{plain}

\theoremstyle{definition}

\theoremstyle{remark}

\theoremstyle{plain}

\newtheorem*{lem*}{Lemma}

\theoremstyle{remark}

\theoremstyle{plain}

\theoremstyle{plain}

\@ifundefined{showcaptionsetup}{}{%
 \PassOptionsToPackage{caption=false}{subfig}}
\usepackage{subfig}
\makeatother

\usepackage{babel}
\providecommand{\claimname}{Claim}
\providecommand{\definitionname}{Definition}
\providecommand{\lemmaname}{Lemma}
\providecommand{\remarkname}{Remark}
\providecommand{\theoremname}{Theorem}
\providecommand{\corollaryname}{Corollary}
\providecommand{\propositionname}{Proposition}

\newcommand{\RN}{\mathbb{R}^N}
\newcommand{\CN}{\mathbb{C}^N}

\newcommand{\diag}{\mathrm{diag}}
\newcommand{\Ci}{C}
\newcommand{\ph}{\mathrm{phase}}

\newcommand{\DFT}[1]{\operatorname{DFT}\!\left( {#1} \right)}

\begin{document}
\title{A Spectral Method for Stable Bispectrum Inversion with Application to Multireference Alignment}

\author{Hua Chen$^\dagger$, Mona Zehni$^\dagger$, and Zhizhen Zhao~\thanks{$\dagger$ These two authors contributed equally to the work. \newline HC, MZ, and ZZ are with the University of Illinois at Urbana-Champaign in the Department of Electrical and Computer Engineering. {The authors were partially supported by UIUC COE SRI grant.}}}

\maketitle

\begin{abstract}
We focus on an alignment-free method to estimate the underlying signal from a large number of noisy randomly shifted observations. Specifically, we estimate the mean, power spectrum, and bispectrum of the signal from the observations. Since bispectrum contains the phase information of the signal, reliable algorithms for bispectrum inversion is useful in many applications. We propose a new algorithm using spectral decomposition of the normalized bispectrum matrix for this task. For clean signals, we show that the eigenvectors of the normalized bispectrum matrix correspond to the true phases of the signal and its shifted copies. In addition, the spectral method is robust to noise. It can be used as a stable and efficient initialization technique for local non-convex optimization for bispectrum inversion. 
\end{abstract}


\section{Introduction}
We consider the problem of estimating a discrete signal with the following observation model,
\begin{equation}
\label{eq:obs_model}
\xi_j = R_{s_j} x + \varepsilon_j, \quad j \in \{1,2,...,M\}
\end{equation}
where $\xi_j \in \mathbb{R}^N$ and $x \in \mathbb{R}^N$, correspond to the $j$-th observation and the underlying signal respectively. $R_{s}: \mathbb{R}^N \rightarrow \mathbb{R}^N$ denotes a cyclic shift operator that shifts the underlying signal, i.e. $(R_s x)[n] = x[n+s \, \textrm{mod} \, N]$. For the sake of brief notations, all indices are understood modulo $N$, namely, in the range $0,\ldots,N-1$. $\varepsilon_j$ represents additive white Gaussian noise with zero mean and unit variance. In \textit{multi-reference alignment (MRA)} \cite{bandeira2014multireference}, both $x$ and the translations $\{s_j\}$ are unknown and the primary goal is to recover $x$ from noisy observations $\{\xi_j\}_{j=1}^M$. 
The alignment-free approach in~\cite{Bendory2017bispec} uses the shift invariant features, such as mean, power spectrum, and bispectrum estimated from the data to recover the underlying signal $x$.

Consider the case where the number of observations is much larger than the length of the signal, namely  $M \gg N$. 
In this regime, the invariant features approach has two important advantages over methods that rely on estimating the translations, 1) there will be no need to determine the translations in order to recover the signal hence reducing the computational complexity of the problem, and 2) with high level of noise, given enough samples, it does not suffer from the fundamental limit~\cite{aguerrebere2016fundamental} for the pairwise alignment approach, that depends on the noise variance.


Phase retrieval from the power spectrum is an ill-posed problem and received considerable attention in recent years~\cite{fienup1982phase,shechtman2015phase,jaganathan2013sparse,bendory2016non,beinert2015ambiguities}. 
Exploiting bispectrum as another invariant feature besides the power spectrum adds the phase information for estimating $x$. The bispectrum  also plays a central role in a variety of signal processing applications. For example, it is a key tool to separate Gaussian and non-Gaussian processes  \cite{brockett1988bispectral,bartolo2004non}. It is also used in seismic signal processing \cite{matsuoka1984phase}, image deblurring \cite{chang1991blur}, MIMO systems~\cite{ChenPet2001}, feature extraction for radar \cite{chen2008feature}, analysis of EEG signals \cite{ning1989bispectral}, cryo-EM image classification~\cite{zhao2014rotationally} and cosmic background radiation analysis \cite{luo1993angular,wang2000cosmic}. A more general setting of bispectrum over compact groups was considered in~\cite{kondor2007novel,kakarala2009completeness,kakarala2012bispectrum,kakarala2009bispectrum}. Many of those applications require reliable algorithm to invert the bispectrum. 

A variety of algorithms have been proposed to invert bispectrum~\cite{giannakis1989signal,sadler1992shift,matsuoka1984phase, Petropulu1998, Bendory2017bispec}. 
In particular, two non-convex optimization algorithms on the manifold of phases in~\cite{Bendory2017bispec} are able to recover phases exactly (up to global time shift) with random initialization in the absence of noise. Several initialization algorithms were also proposed in \cite{Bendory2017bispec}, such as semidefinite programming (SDP), phase unwrapping by integer programming (IP) and frequency marching (FM). The SDP and IP methods are computationally expensive, and therefore do not scale well with the signal length. Although FM is efficient with $O(N^2)$ computational complexity, it is not robust to noise and can be unstable. 

In this letter, we propose a new initialization algorithm using spectral decomposition of the normalized bispectrum matrix. For clean signals, we show that the eigenvectors of the normalized bispectrum matrix correspond to the true phases of the signal and its shifted copies. With bispectrum estimated from the noisy observations, we propose a rule to select the eigenvector for robust estimation of the phase. The computational complexity of the spectral initialization given the invariant features  
is $O(N^3)$, which is dominated by the eigendecomposition of an $N \times N$ normalized bispectrum matrix.
We compare our spectral method with different bispectral inversion techniques through the relative error of reconstruction and computation time. In addition, we demonstrate that with the spectral initialization, the local non-convex approaches converge much faster than the random initialization. 

The letter is organized as follows. Section~\ref{sec:inv} discusses the alignment-free approach for MRA using invariant features. Section~\ref{sec:spectral} presents the spectral method for bispectrum inversion. Section~\ref{sec:results} is devoted to the numerical experiments that compare several bispectrum inversion methods. 

Throughout the letter we use the following notation. Vectors {$x\in \RN$ and $y\in\CN$} denote the underlying signal and its discrete Fourier transform (DFT), respectively. We use $\tilde{a}= \ph(a)$ for the phase of a complex number and $\overline{a}$ for its complex conjugate. When $a = 0$, we assume its phase $\tilde{a}$ is 1. We denote by $z^*$ the conjugate-transpose of $z$ 
 and denote by $'\circ'$ the Hadamard (entry-wise) product. $\Ci(z)$ denotes circulant matrices determined by their first row $z$, i.e., $\Ci(z)[k_1,k_2]=z[k_2-k_1]$. 

Reproducible research: our code is available in \url{https://github.com/ARKEYTECT/Bispectrum_Inversion}.


\section{Invariant Features for Multireference Alignment }
\label{sec:inv}
We solve the MRA problem directly using features that are invariant under translations, namely, the DC component, power spectrum, and bispectrum of the signal. The DFT of a signal is defined as, $\DFT{x}[k] = \sum\limits_{n=0}^{d-1} x[n] e^{\frac{- 2 \pi i n k}{N}}$. 
Shifting the signal in time domain introduces a phase shift in the Fourier coefficients, $\DFT{R_s x}[k]  = y[k] \cdot e^{2\pi i k s / N}$. The DC component of the signal $y[0] = N \mu_x = \sum\limits_{n=0}^{N-1} x[n]$, where $\mu_x$ is the mean of the signal, is clearly invariant to translation. The distribution of the mean of the observations $\xi_j$ is then given by $\mu_{\xi_j}\sim \mathcal{N}\left(\mu_x,\frac{\sigma^2}{N}\right)$ and the unbiased estimate of $\mu_x$ is defined as
\begin{equation} \label{eq:est_mean_x}
\hat{\mu} = \frac{1}{M}\sum_{j=1}^M \left(\frac{1}{N}\sum_{n=0}^{N-1}\xi_j[n]\right)\sim\mathcal{N}\left({\mu}_x,\frac{\sigma^2}{NM}\right).
\end{equation}
Therefore, the estimated DC component is $\hat{y}[0] = N \hat{\mu}$. 

In addition to the DC term, power spectrum, which is defined as $P_x[k]  = \vert y[k]\vert^2$ for all $k \in \{ 0,1,...,N-1\}$ provides the information of the Fourier magnitudes. Since shifting the signal only affects the phase of the Fourier coefficients, power spectrum is invariant to the translations. An estimator for the power spectrum from noisy samples is 
\begin{align}
	{\hat{P}}[k] & = \frac{1}{M}\sum_{j=1}^M P_{\xi_j}[k]-N\sigma^2 \rightarrow P_x[k],\,\text{as}\,M\rightarrow\infty,
	\label{eq:ps_estimator}
\end{align} 
where its variance is dominated by $\frac{N^2 \sigma^4}{M}$ {for large $\sigma$}. An unbiased estimator of $\sigma$ is derived from $\hat{\sigma}^2  = \frac{1}{N} \operatorname{Var}\left( \sum_{n=0}^{N-1} \xi_j[n] \right)_{j=1,\ldots,M}$. In the letter, we assume that $\sigma$ is known. 


The bispectrum is a function of two frequencies $k_1,k_2= \{0,\dots,N-1\}$ and is defined as,
\begin{equation} \label{eq:bispec}
B_x[k_1,k_2]= y [k_1] \overline{y [k_2]} y[k_2-k_1].
\end{equation}
For any shift $s$,
\begin{align*}
	B_{R_s x}[k_1,k_2] & = \left( y [k_1]e^{2\pi i k_1 s/N}\right)\left( \overline{ y[k_2]}e^{-2\pi i k_2 s /N}\right) \\
	& \quad \cdot \left( y[k_2 - k_1]e^{2\pi i (k_2-k_1) s /N}\right) \\
	& = B_{x}[k_1,k_2]. 
\end{align*}
Hence, the bispectrum is shift invariant and it contains the phase information of the signal $x$.  In matrix notation, we express $B_x$ as
\begin{align}
	B_x = yy^* \circ \Ci(y) = Y \circ \Ci(y),
	\label{eq:bispectrum_matrix}
\end{align}
where $Y = y y^*$ is a rank-one matrix. For the normalized bispectrum $\tilde{B}_x[k_1,k_2]=\ph({B_x}[k_1,k_2])$, $\tilde{B}_x[k_1,k_2] = \tilde{y}[k_1]\overline{\tilde{y}[k_2]}\tilde{y}[k_2-k_1]$.
In matrix notation, it takes the form 
\begin{equation}
\tilde{B}_x = \tilde{Y} \circ \Ci(\tilde{y}), \quad \text{where }\tilde{Y}=\tilde{y}\tilde{y}^*.
\label{eq:normalized_bispec} 
\end{equation}

The expectation of the bispectrum built from noisy observation is 
	\begin{align}
		\mathbb{E} \left\{ B_{\xi_j}[k_1, k_2]\right\} & = B_{x}[k_1, k_2 ] + N \sigma^2 y[0] \left (\delta(k_1, k_2)  \right . \nonumber \\ 
        & \left. \quad \quad + \delta(k_1, 0) + \delta(k_2, 0) \right ).
		\label{eq:bispectrum_estimation}
	\end{align} 
Therefore, by removing the DC component for each $\xi_j$, we get an unbiased estimator
\begin{align} 
	\hat{B} & = \frac{1}{M}\sum_{j=1}^M B_{ \xi_j - \mu_{\xi_j} } \rightarrow  B_{ x - \mu_{x} }, \,\text{as}\,M\rightarrow\infty
	\label{eq:bispec_estimator}
\end{align}
where $\mu_{\xi_j}$ is the mean of signal $\xi_j$, the variance of $\hat{B}$ is controlled by $\frac{N^3 \sigma^6}{M}$ {for large $\sigma$} and the estimator in~\eqref{eq:bispec_estimator} is asymptotically unbiased. Note that to ensure stable estimations of $\mu_x$, $P_x$ and $B_x$, $M$ is required to scale $O(\sigma^2)$, $O(\sigma^4)$ and $O(\sigma^6)$ respectively. 




The signal reconstruction process could be split into three parts, namely, the DC component estimation, Fourier magnitude estimation, and phase estimation. 
Then, after combining the recovered results for $\tilde{y}$ with the mean and power spectrum estimates, an estimate of $y$ denoted by $\hat{y}$ is achieved.  
The estimated real signal $\hat{x}$ is the inverse DFT of $\hat{y}$. Note that the overall complexity of this approach can be relatively low for large number of observations. The computational complexity of deriving estimations of the invariants is $O(M N^2)$ and the storage requirement is $O(N^2)$~\cite{Bendory2017bispec}. Additionally, it requires only one pass over the data which is important when the number of observations is large. The invariant features can be computed in parallel. On the other hand, the computational complexity of the spectral method is $O(N^3)$, leading to an overall complexity of $O(M N^2 + N^3) = O(M N^2)$ for our approach.

In the following section, we further introduce our spectral initialization method that can provide robust and stable phase recovery.

\section{Spectral method for phase recovery in bispectral inversion}
\label{sec:spectral}
Since $x$ is real, $y[k]=\overline{y[-k]}$ and $\Ci (y)$ and $B_x$ in ~\eqref{eq:bispectrum_matrix} are Hermitian matrices.
The circulant matrix $\Ci (\tilde{y} )$ can be diagonalized by the normalized DFT matrix $F$ with $F_{ij} = \frac{1}{\sqrt{N}} e^{-\imath \frac{2\pi i j }{N}}$, that is, 
\begin{equation}
\Ci(\tilde{y}) = F \diag(F \tilde{y}) F^*.
\end{equation}
Following Eq.~\eqref{eq:normalized_bispec} for a clean signal,
\begin{align}
\label{eq:B_phase}
\tilde{B} & = \tilde{y}\tilde{y}^*\circ \Ci(\tilde{y})  \nonumber \\
& = \diag(\tilde{y}) F  \text{diag}\left(F\tilde{y}\right) F^* \diag(\tilde{y})^*. 
\end{align}
Since $\tilde{y}[k]$ are phases, the matrix $ V = \diag(\tilde{y}) F $ is a unitary matrix, i.e. $V V^* = V^* V = I$. Therefore, columns of $ V $ are the eigenvectors of $\tilde{B}$ and $F\tilde{y}$ contains the corresponding eigenvalues. We use $S$ to denote the permutation matrix that sorts $F \tilde{y}$ in descending order. $S$ is an orthogonal matrix $SS^\top = S^\top S = I$. Therefore, we can rewrite Eq.~\eqref{eq:B_phase} and, 
\begin{align}
\label{eq:B_phase2}
\tilde{B} & = V S^\top S \diag \left(F \tilde{y} \right ) S^\top S  V^* \nonumber \\
& = V S^\top \diag \left( S F\tilde{y} \right ) (V S^\top)^*  \nonumber \\
& = U \Lambda U^*,  
\end{align} 
where the diagonal matrix $\Lambda$ contains the eigenvalues of $\tilde{B}$ sorted in descending order, $\lambda_1 \geq \lambda_2 \dots \geq \lambda_N$ with the corresponding eigenvectors $U = [u_1, u_2, \dots u_N]$. Eq. \eqref{eq:B_phase2} reveals that $\tilde{y}$ is encoded in both $\Lambda$ and $U$. If the order of the eigenvalues in $\Lambda$ happen to be exactly the same as $F \tilde{y}$, i.e. $S = I $, then the phases can be computed from the inverse DFT of the sorted eigenvalues. 

However, in general, as the eigenvalues are sorted in a descending fashion in $\Lambda$, they correspond to an unknown permutation of $F \tilde{y}$. Recovering $\tilde{y}$ from $\Lambda$ involves searching through all possible permutations of the eigenvalues, which is computationally expensive. Thus, we exploit $U$ in order to derive $\tilde{y}$. Each column of $U$ contains phases of the original signal and their shifted versions, $u_i[k] = \frac{1}{\sqrt{N}}\tilde{y}[k] e^{-\imath \frac{2\pi s_i k}{N}}$ where $s_i \in [0,\dots, N-1]$. 
In the clean case, we can recover $\tilde{y}$ from any column of $U$ with distinct eigenvalues. In fact the signals recovered from columns of $U$ are related through a global cyclic shift. However, in the noisy case, some of the columns of $U$ are deeply contaminated by noise that they contain very little information about $\tilde{y}$, thus we can not choose any random column of $U$.

We propose to select the eigenvector of the normalized bispectrum matrix with the largest minimum spectral gap. 
With $\{\lambda_i\}_{i = 1}^N$ sorted in descending order, 
the minimum spectral gap $\Delta_i$ for $\lambda_i$ is defined as,
\begin{align}
\Delta_i & = \min(\lambda_{i-1} - \lambda_i, \lambda_i - \lambda_{i+1}), \, \text{for } 1 < i < N\nonumber \\
\Delta_1 & = \lambda_1 - \lambda_2, \, \, \text{and } \Delta_N = \lambda_{N-1} - \lambda_N. 
\label{eq:spec_gap}
\end{align}
Let us denote by $v$ the selected eigenvector. We assume the DC component phase is $1$, i.e. $\hat{v}[0] = 1$ therefore we normalize by $\hat{v}[k] = v[k] / v[0]$. Since each entry of $\hat{y}$ should have unit norm, therefore, the phase estimate $\tilde{y}[k] = \ph(\hat{v}[k])$. The steps for our algorithm are illustrated in Algorithm~\ref{alg:spectral}.
If there are multiple eigenvectors with the same spectral gap, we then consider the one with largest absolute eigenvalue.


 
\begin{algorithm}
\textbf{Input:} The normalized bispectrum $\tilde{B}[k_1,k_2]$ \\
\textbf{Output:} ${\hat{y}}$: an estimation of ${\tilde{y}}$ \\
\textbf{Compute:} $\tilde{B} = U \Lambda U^*$, with $\Lambda = \diag(\{\lambda_i\}_{i=1}^N)$ and $\lambda_1 \geq \lambda_2 \geq \dots \geq \lambda_N$.
\begin{enumerate}
\item $v = u_i$ with the largest spectral gap $\Delta_i$  in \eqref{eq:spec_gap}
\item $\hat{v}[k] = v[k]/{v[0]}$
\item $\tilde{y}[k] = {\hat{v}[k]}/{|\hat{v}[k]|} $
\end{enumerate}
\protect\caption{Spectral Method for Bispectral Phase Retrieval}
\label{alg:spectral}
\end{algorithm}

\section{Numerical Experiments}
\label{sec:results}
The experiments were conducted as follows. Code for proposed algorithms is available online. The true signal $x$ of length $N = 41$ is a standard Gaussian random signal. $M$ randomly shifted and noisy copies of $x$ are generated following ~\eqref{eq:obs_model}. We assess our method through relative reconstruction error which is defined as,
\begin{equation*}
\textrm{relative error}(x,\hat x)= \min_{s\in\{0,\dots,N-1\}}\frac{\|R_{s}\hat{x}-x\|_2}{\|x\|_2},
\end{equation*}

We compare our spectral method with several algorithms described in \cite{Bendory2017bispec} such as iterative phase synchronization, optimization on phase manifold, frequency marching, phase unwrapping, SDP relaxation and known-shifts oracle.
The oracle can align the underlying signals perfectly and the aligned signals are averaged. Here the estimation error is only due to the noise and not mis-alignment. This oracle is meant to capture the best possible performance. Note that for the Riemannian trust-region method~\cite{genrtr} required for the non-convex algorithm on the manifolds of phases, we used Manopt toolbox~\cite{manopt}. The tool box is also applied for phase synchronization algorithms for $15$ iterations with warm-start. All results are averaged over $50$ repetitions. In the sequel, we further discuss the effect of various parameters on the performance of our method in comparison to the other algorithms. Following experiments are conducted on computer with Intel i7 7th generation quad core CPU. 
 
$\bullet$ \textit{Effect of noise level ($\sigma$) on relative reconstruction error and computation time}
 
\begin{figure}
  \centering
  \includegraphics[width=0.4\textwidth]{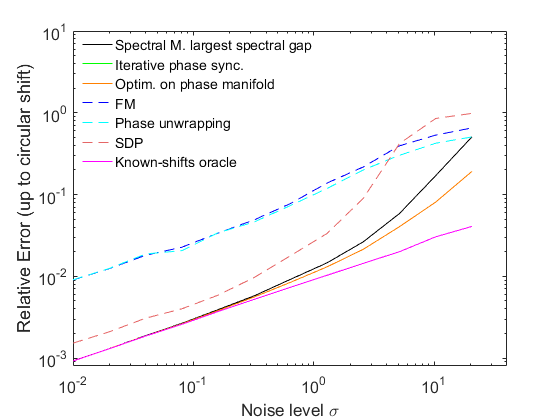}
  \caption{Relative reconstruction error for the signal $x$ as a function of the noise standard deviation $\sigma$ with $M = 10^4$ copies of observations. Note that curves regarding the optim. phase manifold and the iter. phase synch. overlap.}
  \label{fig:rec_error1}
\end{figure}

\begin{figure}
  \centering
  \includegraphics[width=0.4\textwidth]{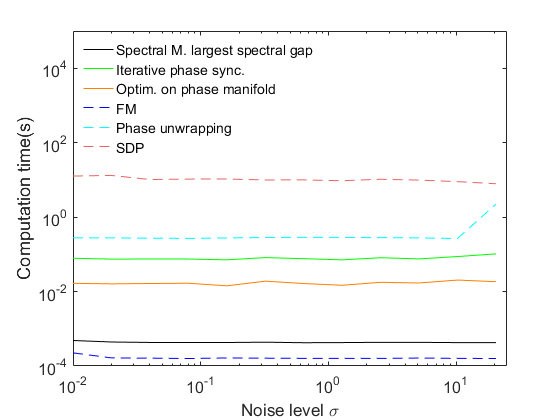}
  \caption{Average computation time over 50 iterations regarding to Fig.~\ref{fig:rec_error1}}
  \label{fig:compute_time1}
\end{figure}

Figures~\ref{fig:rec_error1} and~\ref{fig:compute_time1} show relative reconstruction error as well as computation time of all approaches mentioned above as a function of noise level $\sigma$. Known-shifts oracle provides the most accurate reconstructed signal, followed by iterative phase synchronization, and optimization on phase manifold. 
For $\sigma \leq 0.32$, spectral method is comparable to optimization on phase manifold and iterative phase synchronization in terms of relative reconstruction error. Also, Figure~\ref{fig:compute_time1} verifies the computational efficiency of the spectral method compared to other algorithms. 

$\bullet$ \textit{Effect of the number of observations ($M$) on the relative reconstruction error}

\begin{figure}
  \centering
  \includegraphics[width=0.4\textwidth]{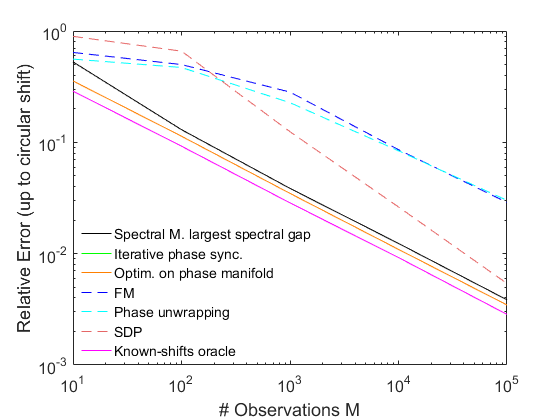}
  \caption{Relative Reconstruction error for the signal $x$ as a function of numbers of observations $M$. Note that curves regarding the optim. phase manifold and the iter. phase synch. overlap.}
  \label{fig:rec_error2}
\end{figure}

\begin{figure}
  \centering
  \includegraphics[width=0.4\textwidth]{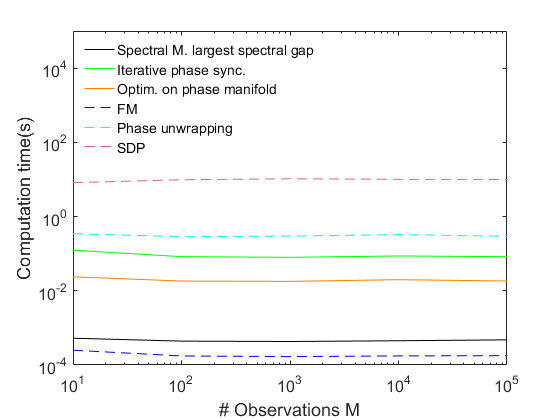}
  \caption{Average computation time over 50 iterations regarding to Fig.~\ref{fig:rec_error2}}
  \label{fig:compute_time2}
\end{figure}

We fix the noise level at $\sigma = 1$ and vary the number of observations from $10$ to $10^5$. Figure~\ref{fig:rec_error2} presents the recovery reconstruction error of the spectral method with respect to the optimization on phase manifold method and iterative phase synchronization as a function of the number of observations $M$. The known-shift oracle indicates the lower bound for the recovery. The performance of two non-convex optimization methods are almost identical and they outperform the spectral method in terms of reconstruction error (see Figure~\ref{fig:rec_error2}). Spectral method is computationally more efficient (see Figure~\ref{fig:compute_time2}).  
\begin{figure}[ht!]
  \centering
  \includegraphics[width=0.4\textwidth]{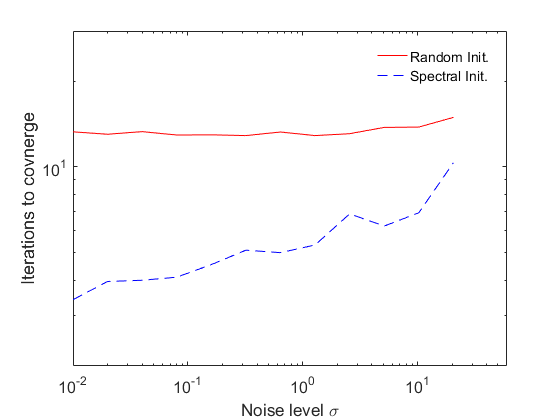}
  \caption{Average number of iterations required to converge for Optimization on Phase Manifold over 50 experiments}
  \label{fig:Niter_opt}
\end{figure}

$\bullet$ \textit{Effect of the spectral initialization on the convergence rate of MRA optimization-based methods} 

Figures~\ref{fig:Niter_opt} shows how initialization with the spectral method changes the number of required iterations for the optimization on phase manifold approach (using Reimannian trust region method) to converge. We initialize $\hat{z}$ with our spectral estimate. We observe that the spectral initialization significantly reduces the number of iterations required for convergence. 
Additionally, as $\sigma$ increases, the number of required iterations in order to obtain convergence rises, which is because the quality of the spectral initialization degrades as the noise level rises. 


\section{Conclusion}
In this letter, we introduced a new spectral initialization approach for recovering a signal from its noisy randomly shifted copies. 
The random shifts are unknown. Instead of trying to recover the relative translations and estimate the denoised signal by averaging the aligned signal, we use shift invariant features to estimate the underlying signal. The invariant features approach has low computational complexity for large sample size compared to alternative methods, such as maximum marginalized likelihood~\cite{Bendory2017bispec}. Previously, two non-convex optimization approaches were proposed along with several initialization algorithms. The new spectral method for bispectral inversion achieves exact recovery when the observations are clean and is robust to noise. It also significantly reduces the number of iterations required for convergence of non-convex optimization methods 
compared to random initialization when the observations are noisy. 

\section*{Acknowledgement}
The authors would like to thank Nicolas Boumal, Tamir Bendory, and Amit Singer for valuable discussions. 

\bibliographystyle{IEEEbib}
\bibliography{ref}

\end{document}